\documentclass[a4paper]{jpconf}

\usepackage{graphicx}
\usepackage{latexsym,amssymb,amsmath,amsfonts}  

\usepackage{dsfont}                   
\newcommand{\openoneFRK}{\mathds{1}}  

\begin{document}

\title{On a soliton-type spacetime defect\,\footnote{Invited talk  
at the \textit{Ninth International Workshop DICE2018:
Spacetime -- Matter -- Quantum Mechanics},
Castiglioncello, Tuscany, Italy, September 17--21, 2018.
\hfill (arXiv:1811.01078)
}
}

\author{F R Klinkhamer}
\address{Institute for Theoretical Physics,
Karlsruhe Institute of Technology (KIT), 76128 Karlsruhe, Germany}

\ead{frans.klinkhamer@kit.edu}

\begin{abstract}
We review the construction of a particular
soliton-type solution of the classical
Einstein and matter-field equations.
This localized finite-energy static classical solution can be interpreted
as a single spacetime defect embedded in Minkowski spacetime
and may give rise to several new effects.
For a Skyrme-type theory with small enough matter-field energy scale
compared to the Planck energy scale
and for a sufficiently small defect length scale,
the existence of a globally regular
solution requires a negative active gravitational mass,
so that the defect repels a distant test particle  (``antigravity'').
There also exist ``stealth defects'' which
have a vanishing asymptotic gravitational mass.
These stealth defects are, however, not entirely invisible as they
bring about a new type of gravitational lensing.
\end{abstract}

\thispagestyle{plain}\pagenumbering{arabic}  

\section{Introduction}
\label{sec:Introduction}

One hypothesis is  that the Universe started out
in some form of ``quantum phase''
which gave rise to classical spacetime and gravity,
as described by Einstein's General Theory of
Relativity~\cite{MisnerThorneWheeler2017}.
It is then possible that this process is analogous
to the cooling of a liquid, which produces an atomic crystal.
But, if the cooling of the latter process is rapid,
the resulting crystal will be imperfect, containing
crystallographic defects.
For the above-mentioned quantum phase and the resulting
classical spacetime, the analogy suggests the possibility
of having ``spacetime defects''
(i.e., imperfections in the fabric of spacetime).
Remark that, historically,
some of the earliest ideas on a foam-like structure of spacetime
go back to Wheeler in the 1950s
(cf. Sec.~43.4 and Box~44.3 in Ref.~\cite{MisnerThorneWheeler2017}
and Chap.~6 in Ref.~\cite{Visser1995}).

Little is known for sure
about the quantum phase of spacetime.
Loop quantum gravity, for example, does have something like ``atoms of space,''
but the emergence of a classical spacetime
has not been established
(cf. Refs.~\cite{AshtekarLewandowski1997a,AshtekarLewandowski1997b,%
Rovelli2008}  and references therein).

Here, we will stay on the classical side of the problem
and use the framework of Einstein's General Relativity (GR).
Specifically, we will obtain a soliton-type classical solution
to describe a single spacetime defect and we will
investigate certain novel effects which
this soliton-type solution produces.

The aim of the present contribution is to provide a
more or less self-contained  discussion  
of one particular soliton-type classical solution,
namely a Skyrmion spacetime defect solution~\cite{Klinkhamer2014-prd}
and to explain the origin of certain new
phenomena~\cite{KlinkhamerQueiruga2018-antigrav,%
KlinkhamerQueiruga2018-stealth,{KlinkhamerWang2019-lensing}}.
In Sec.~\ref{sec:Skyrmion-spacetime-defect},
we first review some background material on the manifold considered
(see also Refs.~\cite{Klinkhamer2014-mpla,%
KlinkhamerSorba2014,Schwarz2010,Guenther2017}),
then define the fields and their action,
and, finally, present the details of the Skyrmion spacetime defect solution.
In Sec.~\ref{sec:Antigravity},
we explain why certain
Skyrmion spacetime defects produce
``antigravity,'' that is, repulsion of a distant test particle.
In Sec.~\ref{sec:Stealth-defect},
we consider other Skyrmion spacetime defects which have
a positive energy density of the matter fields
but a vanishing asymptotic gravitational mass,
so that these defects may be called ``stealth defects.''
In Sec.~\ref{sec:Lensing},
we show that such stealth defects are not entirely
invisible, as they give lensing of light
(this lensing is different from standard gravitational lensing).
In Sec.~\ref{sec:Discussion}, we, first,
compare our soliton-type spacetime defect
with another kind of spacetime
defect~\cite{Hossenfelder2013,Hossenfelder-etal2018}
and, then, review the general properties of
the Skyrmion spacetime defect solution.

Before we start with the technical discussion of the next section,
we emphasize that our Skyrmion spacetime defect is a genuine
solution of standard GR, as long as we allow for degenerate metrics.
The hope is that the equations of GR know about the
``edge of the theory'' and that they may provide
some insight into the nature of possible defects of spacetime.

\section{Skyrmion spacetime defect}
\label{sec:Skyrmion-spacetime-defect}

\subsection{Basic idea}
\label{subsec:Basic-idea}

\begin{figure}[h]  
\begin{center}
\includegraphics[width=0.45\textwidth]{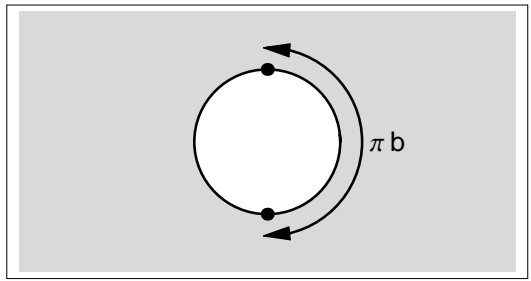}
\end{center}
\vspace*{-2mm}
\caption{Three-space $\widetilde{M}_{3}$  obtained by surgery
from the Euclidean 3-space $E_3\,$: interior of a ball with radius $b$
removed and antipodal points on the boundary of the ball
identified (as indicated by the dots).
The ``long distance'' between identified points on the
defect surface equals $\pi \, b$.}
\label{fig1}
\end{figure}

Let us start by presenting a rough description of our
spacetime defect solution~\cite{Klinkhamer2014-prd},
with mathe\-matical details to follow later.
The basic idea is to
obtain a \emph{nonsingular} finite-energy static defect solution
of the Einstein field equation, with a length parameter
$b>0$ and topology as suggested by the sketch in
Fig.~\ref{fig1}.

An important ingredient of such a regular solution
is the choice of \emph{appropriate coordinates}.
The standard Cartesian
coordinates of Euclidean 3-space are unsatisfactory,
as a single point
may have different coordinates.
Taking the origin of the Cartesian coordinate system 
to co\-in\-cide with the center of the 3-ball
in Fig.~\ref{fig1}, the coordinates
$(x^1,\,  x^{2},\, x^3)=(0,\,b,\,0)$ and
$(x^1,\,  x^{2},\, x^3)=(0,\,-b,\,0)$, for example,
correspond to the same point (the identified dots in Fig.~\ref{fig1}).
Instead of a single chart of Cartesian coordinates,
it is possible to use three overlapping charts of coordinates,
each one centered on one
of the three Cartesian coordinate axes~\cite{Schwarz2010,Guenther2017}.
The promised mathematical details now follow.

\subsection{Manifold}
\label{subsec:Manifold}

The four-dimensional spacetime manifold considered has topology
\begin{equation}\label{eq:M4}
\widetilde{M}_4 = \mathbb{R} \times \widetilde{M}_{3}\,,
\end{equation}
where $\widetilde{M}_{3}$ is a noncompact,
orientable, nonsimply-connected manifold without boundary.
Up to a point, $\widetilde{M}_{3}$ is homeomorphic
to the 3-dimensional real-projective plane,
\begin{equation}\label{eq:M3-topology}
\widetilde{M}_{3} \simeq
\mathbb{R}P^3 -  p_{\infty}\,,
\end{equation}
where $p_{\infty}$ corresponds to the ``point at spatial infinity.''

For the explicit construction of $\widetilde{M}_{3}$,
we perform \emph{local surgery} (Fig.~\ref{fig1})
on  the 3-dimensional Euclidean space
$E_{3}=\big(\mathbb{R}^3,\, \delta_{mn}\big)$.
Recall the standard Cartesian and spherical
coordinates \mbox{on $\mathbb{R}^3$},
\begin{subequations}\label{eq:Cartesian-spherical-coord-defs-ranges}
\begin{eqnarray}\label{eq:Cartesian-spherical-coord-defs}
\vec{x}
&\equiv&
|\vec{x}|\, \widehat{x}
= (x^1,\,  x^{2},\, x^3)
= (r \sin\theta  \cos\phi,\,r \sin\theta  \sin\phi,\, r \cos\theta )\,,
\end{eqnarray}
with ranges
\begin{eqnarray}\label{eq:Cartesian-spherical-coord-ranges}
x^m &\in& (-\infty,\,+\infty)\,,
\quad
r \geq 0\,,
\quad
\theta \in [0,\,\pi]\,,
\quad
\phi \in [0,\,2\pi)\,.
\end{eqnarray}
\end{subequations}
Now, $\widetilde{M}_{3}$ is obtained from $\mathbb{R}^3$ by removing
the interior of the ball $B_b$ with radius $b$ and
identify\-ing antipodal points on its boundary $S_b \equiv \partial B_b$.
With point reflection denoted $P(\vec{x})=-\vec{x}$,
the \mbox{3-space} $\widetilde{M}_{3}$ is given by
\begin{equation}\label{eq:M3-definition}
\widetilde{M}_{3} =
\Big\{  \vec{x}\in \mathbb{R}^3\,:\; \Big(|\vec{x}| \geq b >0\Big)
\wedge
\Big(P(\vec{x})\;\widehat{=}\;\vec{x} \;\;\mathsf{for}\;\;  |\vec{x}|=b\Big)
\Big\}\,,
\end{equation}
where ``$\,\widehat{=}\,$'' stands for point-wise identification.

\begin{figure}[h] 
\begin{center}
\includegraphics[width=0.6\textwidth]{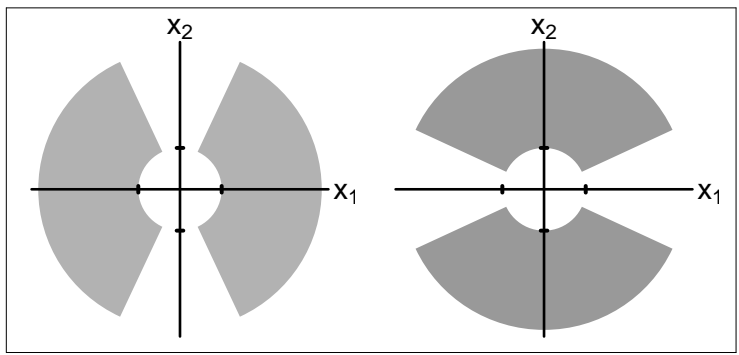}
\caption{Slice $x_{3} = 0$ of the manifold $\widetilde{\mathcal{M}}_{3}$
with the domains of the chart-1 coordinates (left)
and the chart-2 coordinates (right). The tick marks on the
$x_1$ and $x_2$ axes correspond to the values $\pm b$.
The 3-dimensional domains are obtained by revolution
around the $x_1$-axis (left) or the $x_2$-axis (right).
The domain of the chart-3 coordinates is defined similarly.}
\label{fig2}
\end{center}
\end{figure}

As mentioned before, a
relatively simple covering of $\widetilde{M}_{3}$ uses three charts of
coordinates, labeled by $n=1,2,3$.
Each  chart surrounds one of the three Cartesian
coordinate axes~\cite{Schwarz2010,Guenther2017}, 
as illustrated by Fig.~\ref{fig2}.

These coordinate charts are denoted
\begin{equation}\label{eq:XnYnZn}
(X_n,\,  Y_n,\, Z_n)\,, \quad\mathsf{for}\quad n=1,\,2,\,3\,.
\end{equation}
Note that the triples (\ref{eq:XnYnZn})
are, despite appearances, non-Cartesian coordinates.
Specifically, the set of coordinates surrounding the $x^{2}$-axis segments
with $|x^{2}|\geq b$ is given by
(cf. Sec.~II D in Ref.~\cite{KlinkhamerSorba2014})
\begin{subequations}\label{eq:X2Y2Z2-def}
\begin{eqnarray}
X_2 &=& \left\{\begin{array}{ll}
                \phi\,,\phantom{\sqrt{r^{2}-b^{2}}}
                &\;\quad\mathsf{for}\quad  0 < \phi< \pi\,,\\
                \phi-\pi\,,
                &\;\quad\mathsf{for}\quad  \pi < \phi < 2\pi\,,
\end{array}\right.
\\[2mm]
Y_2 &=& \left\{\begin{array}{ll}
                +\sqrt{r^{2}-b^{2}}\,,
                &\;\quad\mathsf{for}\quad  0 < \phi< \pi\,,\\[2mm]
                -\sqrt{r^{2}-b^{2}}\,,
                &\;\quad\mathsf{for}\quad  \pi < \phi < 2\pi\,,
              \end{array}\right.
\\[2mm]
Z_2 &=& \left\{\begin{array}{ll}
                \theta\,,\phantom{\sqrt{r^{2}-b^{2}}}
                &\;\quad\mathsf{for}\quad  0 < \phi< \pi\,,\\
                \pi-\theta\,,
                &\;\quad\mathsf{for}\quad  \pi < \phi < 2\pi\,,
              \end{array}\right.
\end{eqnarray}
with ranges
\begin{eqnarray}\label{eq:X2Y2Z2-ranges}
X_2 \in (0,\,\pi)\,,
\quad
Y_2 \in (-\infty,\,\infty)\,,
\quad
Z_2 \in (0,\,\pi)\,.
\end{eqnarray}
\end{subequations}
The two  other coordinate sets, $(X_1,\,  Y_1,\, Z_1)$
and $(X_{3},\,  Y_{3},\, Z_{3})$, are defined similarly.

In the following, we explicitly discuss only one coordinate chart,
which we take to be (\ref{eq:X2Y2Z2-def}).
Furthermore, the notation is simplified as follows:
\begin{equation}\label{eq:XYZT-def}
(T,\,X,\,  Y,\, Z) \equiv (T,\,X_2,\,  Y_2,\, Z_2)\,,
\end{equation}
where the time coordinate $T$ has been added in order
to describe the spacetime manifold $\widetilde{M}_4$.

\subsection{Fields and action}
\label{subsec:Fields-and-action}

Consider a Skyrme-type scalar field $\Omega(X)\in SO(3)$,
which propagates
over the spacetime manifold (\ref{eq:M4})
and has the following action ($c=\hbar=1$):%
\begin{subequations}\label{eq:action-Lgrav-Lmat-omegamu}
\begin{eqnarray}\label{eq:action}
\hspace*{-0mm}
S &=&
 \int_{\widetilde{M}_4} d^4X\,\sqrt{-g}\,
\Big(\mathcal{L}_{\mathsf{grav}}+ \mathcal{L}_{\mathsf{mat}}\Big)\,,
\\[2mm]
\label{eq:Lgrav}
\hspace*{-0mm}
\mathcal{L}_{\mathsf{grav}}&=& \frac{1}{16\pi\, G_{N}}\:R\,,
\\[2mm]
\hspace*{-0mm}
\label{eq:Lmat}
\mathcal{L}_{\mathsf{mat}}&=&
\frac{f^{2}}{4}\:\mathsf{tr}\Big(\omega_\mu\,\omega^\mu\Big)
+\frac{1}{16\, e^{2}}\: \mathsf{tr}\Big(\left[\omega_\mu,\,\omega_\nu\right]
\left[\omega^\mu,\,\omega^\nu\right]\Big)
+	\frac{1}{2}\,m^{2}\, f^{2}\,\mathsf{tr}\Big(\Omega-\openoneFRK_{3}\Big)\,,
\\[2mm]
\label{eq:omegamu}
\hspace*{-0mm}
\omega_\mu &\equiv& \Omega^{-1}\,\partial_\mu\,\Omega\,,
\end{eqnarray}
\end{subequations}
with $g \equiv \mathsf{det}\, g_{\mu\nu}$  and the Ricci curvature 
scalar $R \equiv R_{\kappa\lambda\mu\nu}\,g^{\kappa\mu}\,g^{\lambda\nu}$. 
Defining ``pions'' $\pi^a$ by
\begin{equation}
\label{eq:Omega-pi-a}
\Omega(X)=\exp\Big[S^a\, \pi^a(X)/f \,\Big]\,,
\end{equation}
for three $3\times 3$ matrices $S^a$ given by
\begin{eqnarray}
\label{eq:S123-def}
S^1 &\equiv&  \left(
                \begin{array}{ccc}
                  0     & 0  &   0 \\
                  0     & 0  &   1 \\
                  \;0\; & -1 & \;0\; \\
                \end{array}
              \right)\,,
\quad
S^{2} \equiv  \left(
                \begin{array}{ccc}
                  \;0\; & \;0\; & -1 \\
                    0   &   0   & 0 \\
                    1   &   0   & 0 \\
                \end{array}
              \right)\,,
              \quad
S^3 \equiv  \left(
                \begin{array}{ccc}
                  0  &   1   &   0 \\
                  -1 & \;0\; & \;0\; \\
                  0  &   0   &   0 \\
                \end{array}
              \right)\,,
\end{eqnarray}
we have
\begin{equation}
\mathcal{L}_{\mathsf{mat}}=
-\frac{1}{2}\,\partial_\mu \pi^a\partial^\mu \pi^a
-\frac{1}{2}\,m^{2}\,\pi^a\pi^a+ \cdots \,.
\end{equation}

The theory \eqref{eq:action-Lgrav-Lmat-omegamu}
has three dimensional parameters and a single dimensionless parameter $e$,
\begin{subequations}\label{eq:parameters-GN-f-m-Skyrme-e}
\begin{eqnarray}
G_{N}&\geq& 0\,,
\\[2mm]
f &>& 0 \,,
\\[2mm]
m &\geq& 0\,,
\\[2mm]
\label{eq:Skyrme-e}
e &>&  0\,.
\end{eqnarray}
\end{subequations}
From these parameters, we obtain the following two dimensionless parameters:
\begin{subequations}\label{eq:etatilde-mtildesquare}
\begin{eqnarray}\label{eq:etatilde}
\widetilde{\eta}&\equiv& 8\pi\, G_{N}\, f^{2} \geq 0\,,
\\[2mm]
\label{eq:mtildesquare}
\widetilde{m} &\equiv& \frac{m}{e\, f} \geq 0 \,.
\end{eqnarray}
\end{subequations}

\subsection{Ans\"{a}tze}
\label{subsec:Ansaetze}

The self-consistent \textit{Ans\"{a}tze} for the metric
and the $SO(3)$ matter field are~\cite{Klinkhamer2014-prd}
\begin{subequations}\label{eq:Ansaetze}
\begin{eqnarray}\label{eq:metric-Ansatz}
\hspace*{-2.0mm}
ds^{2}\,\Big|_{\widetilde{M}_4\,,\,\mathsf{chart-2}}
&=&
 -\big[\mu(W)\big]^{2}\, dT^{2}
 +\big(1-b^{2}/W\big)\,\big[\sigma(W)\big]^{2}\,(d Y)^{2}
\nonumber\\[1mm]&&
+W \,\Big[(d Z)^{2}+\sin^{2} Z\, (d X)^{2} \Big]\,,
\\[2mm]
\label{eq:hedgehog-Ansatz}
\hspace*{-2.0mm}
\Omega(X) &=&
\cos\big[F(W)\big]\;\openoneFRK_{3}
-\sin\big[F(W)\big]\;
\widehat{x}\cdot \vec{S}
+\big(1-\cos\big[F(W)\big]\big)\;
\widehat{x} \otimes \widehat{x}\,,
\\[2mm]
\label{eq:W-definition}
\hspace*{-2.0mm}
W &\equiv& b^{2}+Y^{2}\,,
\end{eqnarray}
\end{subequations}
with a unit 3-vector $\widehat{x} \equiv \vec{x}/|\vec{x}|$
from the Cartesian coordinates $\vec{x}$ defined in terms
of the chart-2 coordinates $X$, $Y$, and $Z$
[see \eqref{eq:Cartesian-spherical-coord-defs-ranges},
\eqref{eq:X2Y2Z2-def}, and \eqref{eq:XYZT-def} above].
Note that the sine-$F$ term in (\ref{eq:hedgehog-Ansatz})
displays the well-known hedgehog behavior, 
linking the ``isospin'' dependance
of the matter field $\Omega(X)$ to its spatial dependance.

The boundary conditions on the three \textit{Ansatz} functions are
\begin{subequations}\label{eq:bcs-F-sigma-mu}
\begin{eqnarray}
\label{eq:bcs-F}
F(b^{2})    &=& \pi\,,\quad F(\infty) = 0 \,,
\\[2mm]
\label{eq:bcs-sigma}
\sigma(b^{2}) &\in&  (0,\,\infty) \,,
\\[2mm]
\label{eq:bcs-mu}
\mu(b^{2})&\in&  (0,\,\infty) \,.
\end{eqnarray}\end{subequations}
The $W=b^{2}$  boundary condition \eqref{eq:bcs-F} 
for \eqref{eq:hedgehog-Ansatz} is consistent
with the topology of $\widetilde{M}_{3}$
[identified antipodal points in Fig.~\ref{fig1}]
and the boundary conditions  (\ref{eq:bcs-sigma})--(\ref{eq:bcs-mu})
will be discussed later.

Two remarks on these \textit{Ans\"{a}tze} are in order.
First, with finite  \textit{Ansatz} functions $\mu(W)$
and $\sigma(W)$, the metric from (\ref{eq:metric-Ansatz})
is \textbf{degenerate} at the defect surface $Y=0$ (or $W=b^{2})\,$,
\begin{equation}
\mathsf{det}\big[g_{\mu\nu}(X)\big]\,\Big|_{W=b^{2}}
=0\,,
\end{equation}
and the standard elementary-flatness property
does not apply (cf. App.~D in Ref.~\cite{Klinkhamer2014-mpla}).
Further discussion on the degeneracy property is relegated to
Sec.~\ref{subsec:Remarks-on-the -Skyrmion-spacetime-defect}.

Second, the matter-field \textit{Ansatz} \eqref{eq:hedgehog-Ansatz}
corresponds to a topologically nontrivial scalar field configuration,
a Skyrmion-like configuration~\cite{Skyrme1961,MantonSutcliffe2004}
with \textbf{unit winding number},
\begin{equation}\label{eq:deg-Omega}
N \equiv \mathsf{deg}
[\Omega]=
-\frac{2}{\pi}\;\int_{\pi}^{0} dF\, \sin^{2}(F/2)=1\,,
\end{equation}
where the endpoints of the integral 
correspond to the boundary conditions (\ref{eq:bcs-F}).
Again, further discussion is relegated to
Sec.~\ref{subsec:Remarks-on-the -Skyrmion-spacetime-defect}.

\subsection{Reduced field equations}
\label{subsec:Reduced-field-equations}

In our numerical analysis, we will use the following dimensionless variables:
\begin{subequations}\label{eq:y-y0-w}
\begin{eqnarray}
\label{eq:y}
y &\equiv& e\,f\;Y \in  \big(-\infty,\,\infty\big)\,,
\\[2mm]
\label{eq:y0}
y_{0} &\equiv& e\,f\;b \in  \big(0,\,\infty\big)\,,
\\[2mm]
\label{eq:w}
w &\equiv& (e\,f)^{2}\;W \equiv y_{0}^{2}+y^{2} \in \big[y_{0}^{2},\,\infty\big)\,,
\end{eqnarray}
\end{subequations}
and the \textit{Ansatz} functions are simply written as
$\sigma(w)$, $\mu(w)$, and $F(w)$.
The reduced field equations are three
ordinary differential equations (ODEs).
With the auxiliary functions
\begin{eqnarray}\label{eq:A-C-def}
A(w) &\equiv& 
2\,\sin^{2}\frac{F(w)}{2} \left(\sin^{2}\frac{F(w)}{2}+w\right)  \,,
\quad
C(w) \equiv  4\,\sin^{2}\frac{F(w)}{2}+w \,,
\end{eqnarray}
these ODEs are~\cite{KlinkhamerQueiruga2018-stealth,Guenther2017}%
\begin{subequations}\label{eq:ODEs-with-pion-mass}
\begin{eqnarray}
\label{eq:ODE-sigma-with-pion-mass}
\hspace*{-0.00mm}
4\,w\,\sigma'(w)
&=&
+\sigma(w)\,\left[
\left[1-\sigma^{2}(w)\right]
+\widetilde{\eta}\, \frac{2}{w}\,
\Big(A(w)\,\sigma^{2}(w) + C(w)\,\left[w\,F'(w)\right]^{2}\Big)
\right]\nonumber\\
&&
+2\, w\,\widetilde{m}^{2}\,\widetilde{\eta}\,
\sigma^3(w)\,
\sin^{2}\frac{F(w)}{2}\,,
\\[2.0mm]
\label{eq:ODE-mu-with-pion-mass}
\hspace*{-0.00mm}
4\,w\,\mu'(w)
&=&
-\mu(w)\,\left[
\left[1-\sigma^{2}(w)\right]
+ \widetilde{\eta}\, \frac{2}{w}\,
\Big(A(w)\,\sigma^{2}(w) - C(w)\,\left[w\,F'(w)\right]^{2}\Big)
\right]\nonumber\\
&&
-2\, w\,\widetilde{m}^{2}\,\widetilde{\eta}\,\sigma^{2}(w)\,\sin^{2}\frac{F(w)}{2}\,,
\\[2.0mm]
\label{eq:ODE-F-with-pion-mass}
\hspace*{-0mm}
C(w)\,w^{2}\,F''(w)
&=&
+\sigma^{2}(w)\,\sin F(w)
\,\left( \sin^{2}\frac{F(w)}{2}+\frac{w}{2}  \right)
-\frac{1}{2}\,C(w)\,\sigma^{2}(w)\,w\,F'(w)
\nonumber\\&&
\times
\,\left[1-4\,\widetilde{\eta}\,\frac{1}{w}\, \sin^{2}\frac{F(w)}{2}
 \,\left(\sin^{2}\frac{F(w)}{2} +w  \right) \right]
\nonumber\\&&
-w\,F'(w)
\,\Big[w\,F'(w)\,\sin F(w)+w \Big]
+\frac{\widetilde{m}^{2}}{2}\,w^{2}\,\sigma^{2}(w)\,\sin\frac{F(w)}{2}
\nonumber\\&&
\times
\left[\cos\frac{F(w)}{2}
+2\,\widetilde{\eta}\,C(w)\,\sin\frac{F(w)}{2}F'(w)\right]\,,
\end{eqnarray}
\end{subequations}
where the prime stands for differentiation with respect to $w$.

\subsection{Matter energy density and gravitational mass}
\label{subsec:Matter-energy-density-and-gravitational-mass}

For later use, we also give the reduced expression for the
00-component of the energy-momentum tensor $T^{\mu}_{\;\;\nu}(w)$
which corresponds to the negative of the energy density $\rho(w)$,
\begin{eqnarray}
\label{eq:T-up0-down0}
T^0_{\;\;0}(w)
&=&
-f^{2}\,(e\,f)^{2}\, \frac{2}{w^{2}\,\sigma^{2}(w)}
\nonumber\\
&&
\times
\Bigg(
  A(w)\,\sigma^{2}(w)
+ C(w)\,\big[w\,F'(w)\big]^{2}
+ \widetilde{m}^{2}\,w^{2}\,\sigma^{2}(w) \,\sin^{2}\frac{F(w)}{2}
\Bigg) \,.
\end{eqnarray}
The expression in the large brackets on the right-hand side
of \eqref{eq:T-up0-down0} has already appeared on the right-hand side
of \eqref{eq:ODE-sigma-with-pion-mass}, which resulted from
the 00-component of the Einstein equation.

A further useful quantity is the dimensionless
length scale $l(w)$, defined by
\begin{equation}\label{eq:def-l}
l(w) \equiv \sqrt{w}\,\left[1-\frac{1}{\sigma^{2}(w)}\right]\,,
\end{equation}
which corresponds to a Schwarzschild-type behavior
of the square of the metric function,
\begin{equation}\label{eq:sigma-from-l}
\sigma^{2}(w)=\frac{1}{1-l(w)/\sqrt{w}\,}\,.
\end{equation}
The $\sigma$--ODE \eqref{eq:ODE-sigma-with-pion-mass} gives immediately
\begin{eqnarray}
\label{eq:l-ODE}
l'(w)  &=&
\widetilde{\eta}\,w^{-3/2}\;
\left(A(w)+C(w)\,\frac{\left[w\,F'(w)\right]^{2}}{\sigma^{2}(w)} \right)\,,
\end{eqnarray}
where the prime stands for the derivative $d/dw$
and the auxiliary functions $A(w)$ and $C(w)$ are given
by \eqref{eq:A-C-def}.
As the right-hand-side of \eqref{eq:l-ODE} is manifestly nonnegative,
the interpretation is that the Schwarzschild-type length scale $l(w)$,
corresponding to a $w$-dependent effective mass,
can only increase by the addition of positive energy density
from the matter fields~\cite{KlinkhamerQueiruga2018-antigrav}.
The asymptotic value of $l(w)$ defines
the Arnowitt--Deser--Misner (ADM) mass in our context,
\begin{equation}\label{eq:M-ADM}
M_{\mathsf{ADM}} =
l_\infty\big/\big(2\, G_{N} \,e\,f\big)\,,
\quad
l_\infty \equiv \lim_{w\rightarrow\infty} l(w) \,,
\end{equation}
which corresponds to the  active gravitational mass
of the soliton-type solution~\cite{Schwarz2010,Guenther2017}.

\subsection{Numerical solutions}
\label{subsec:Numerical-solutions}

The ODEs \eqref{eq:ODEs-with-pion-mass} can be solved numerically with
boundary conditions  from (\ref{eq:bcs-F-sigma-mu}).
Specifically, we have
$F(y_{0}^{2}) = \pi$ and $F(\infty) = 0$
for the matter-field \textit{Ansatz}
function $F(w)$
and we, first, take $\sigma(y_{0}^{2})=1$
for the metric \textit{Ansatz} function $\sigma(w)$
[the value of $\mu(y_{0}^{2})$ can be rescaled arbitrarily].

\begin{figure*}[b]
\vspace*{-5mm}
\begin{center}
\includegraphics[width=0.8\textwidth]{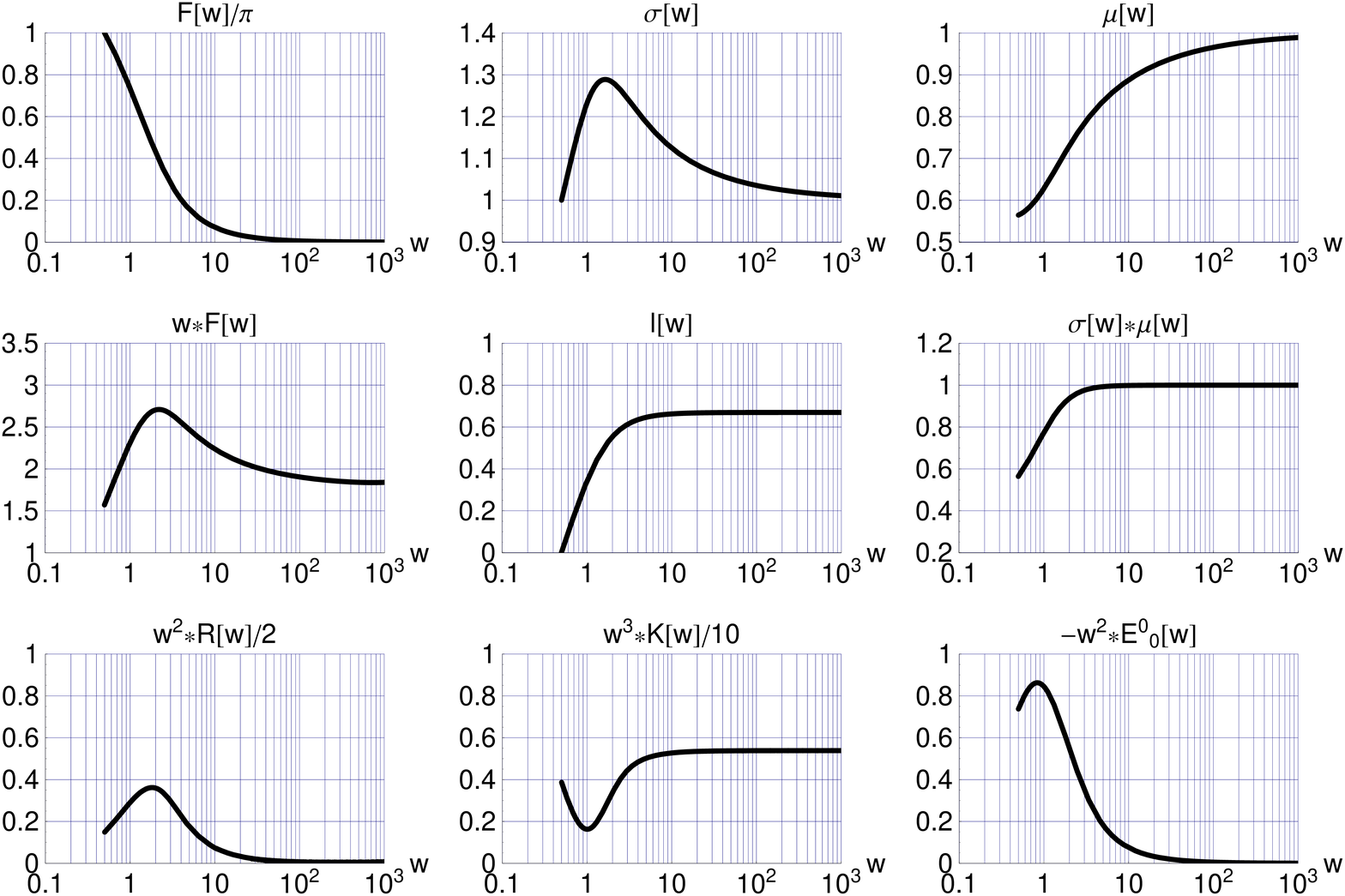}
\end{center}
\vspace*{-5mm}
\caption{Numerical solution~\cite{KlinkhamerQueiruga2018-antigrav}
of the reduced field equations
\eqref{eq:ODEs-with-pion-mass} with parameters
\mbox{$\widetilde{\eta}\equiv 8\pi\, G_{N}\, f^{2} =1/20$},
$\widetilde{m}\equiv m/(e\, f)=0$,
and $y_{0}\equiv e f b=1/\sqrt{2}$.
The boundary conditions at the defect surface
$w=y_{0}^{2}=1/2$ are:
$F=\pi$,
$F^\prime=-1.9718377138$,
$\sigma=1$,
and  $\mu=0.564337$.}
\label{fig3}
\vspace*{-2mm}
\end{figure*}

Figure~\ref{fig3} shows the
\textit{Ansatz} functions $F(w)$, $\sigma(w)$, and $\mu(w)$ of
one particular numerical solution. Also displayed are
the dimensionless Ricci curvature scalar $R(w)$,
the dimensionless Kretschmann curvature  scalar 
$K(w) \equiv  R_{\kappa\lambda\mu\nu}(w)\,R^{\kappa\lambda\mu\nu}(w)$, and
the negative of the
$00$ component of the dimensionless Einstein
tensor $E^{\mu}_{\;\;\nu}(w)$
$\equiv$ $R^{\mu}_{\;\;\nu}(w) -(1/2)\,R(w)\,\delta^{\mu}_{\;\;\nu}$
[note that $-E^{0}_{\;\;0}(w)$ is proportional to the energy
density $\rho(w)$ by the Einstein equation].
This Fig.~\ref{fig3} shows, moreover, the behavior of
the dimensionless
Schwarzschild-type length scale $l(w)$ defined by \eqref{eq:def-l}, which,
as mentioned in Sec.~\ref{subsec:Matter-energy-density-and-gravitational-mass},
stays constant or increases with increasing $w$.
All physical quantities
are well-behaved at the defect surface $w=y_{0}^{2}$, as shown by
Fig.~9 in Ref.~\cite{KlinkhamerQueiruga2018-antigrav}.

The boundary condition $\sigma(y_{0}^{2})=1$
may be called the ``standard'' boundary condition,
because the limit $b\to 0$ then connects to the
standard Minkowski spacetime manifold.
But with $b\ne 0$ and the nontrivial topology
$\mathbb{R}P^3$ from (\ref{eq:M3-topology}),
the boundary condition on the metric \textit{Ansatz} function
can be generalized,
\begin{equation}\label{eq:sigma-bc}
\sigma\big( y_{0}^{2} \big) \in (0,\,\infty)\,,
\end{equation}
where the value zero has been excluded,
in order that the field equations be well-defined
at the $w=y_{0}^{2}$ defect surface~\cite{Guenther2017}.

\begin{figure*}[t]
\begin{center}
\includegraphics[width=0.8\textwidth]{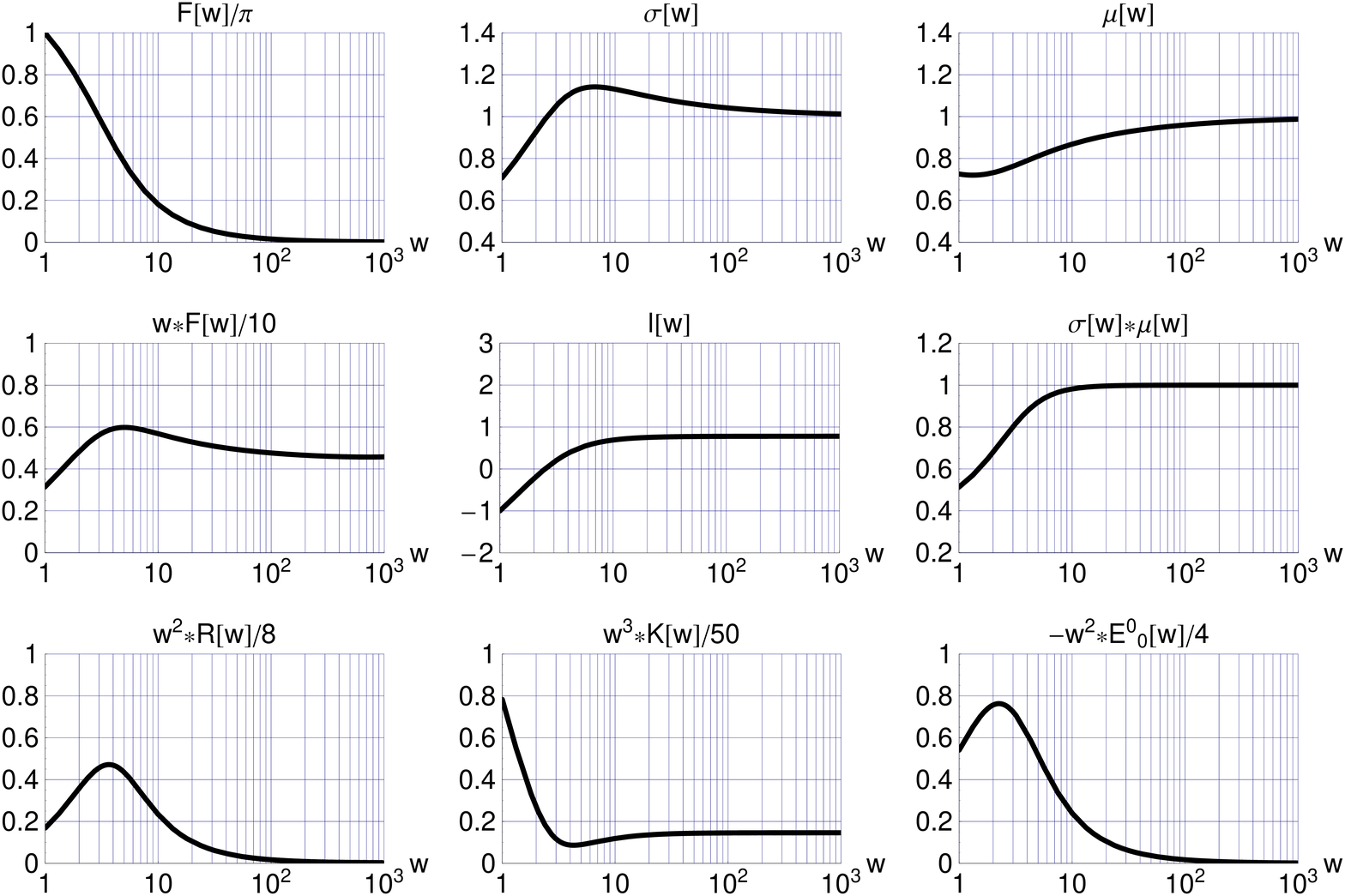}
\end{center}
\caption{Numerical solution~\cite{KlinkhamerQueiruga2018-antigrav}
of the ODEs \eqref{eq:ODEs-with-pion-mass}
with parameters $\widetilde{\eta}=1/10$,
$\widetilde{m}=0$, and $y_{0}=1$.
The boundary conditions at
$w=y_{0}^{2}=1$ are: $F=\pi$, $F^\prime=-0.82561881304$,
$\sigma=1/\sqrt{2}$,
and  $\mu=0.725818$.
This numerical solution with $l_{\infty} \approx 0.8$
has a positive ADM mass \eqref{eq:M-ADM}.}
\label{fig4}
\vspace*{16mm}
\begin{center}
\includegraphics[width=0.8\textwidth]{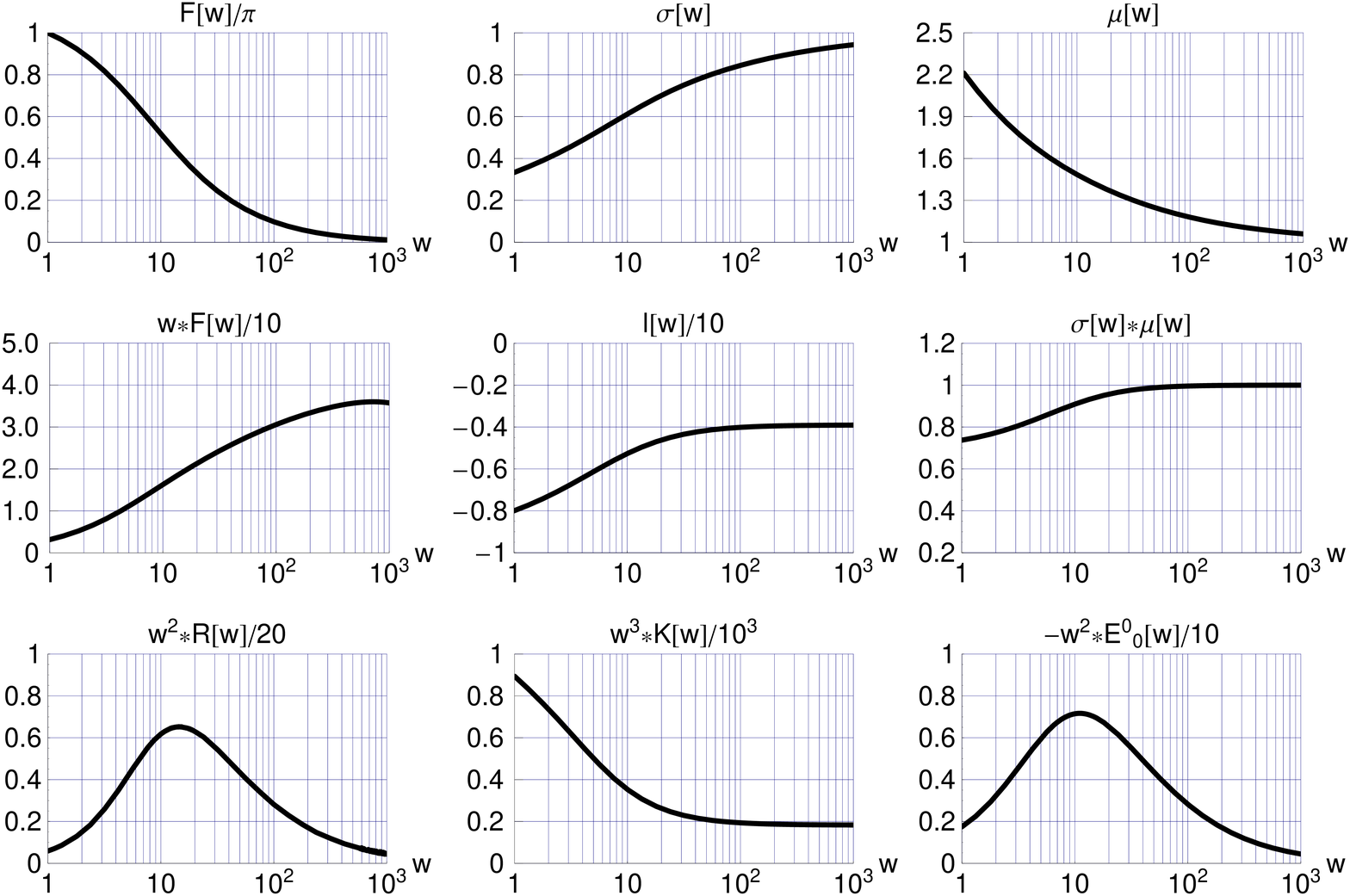}
\end{center}
\caption{Same as Fig.~\ref{fig4}, but now with a smaller value
of $\sigma(y_{0}^{2})$. Specifically, the boundary conditions at
$w=y_{0}^{2}=1$ are:
$F=\pi$, $F^\prime=-0.323978148$,
$\sigma=1/3$,
and  $\mu=2.21176$.
This numerical solution with $l_{\infty} \approx -4$
has a negative ADM mass \eqref{eq:M-ADM}.}
\vspace*{-12mm}  
\label{fig5}
\end{figure*}

Figures~\ref{fig4} and \ref{fig5} give numerical solutions
for two different values of $\sigma(y_{0}^{2})<1$.
The results of Figs.~\ref{fig3}--\ref{fig5} are to be understood
as having \emph{different} values of the
boundary condition $\sigma(y_{0}^{2})$ for \emph{equal}
boundary conditions $F(y_{0}^{2})=\pi$ and  $F(\infty)=0$
[the numerical values for $F'(y_{0}^{2})$ in the figure
captions are mentioned purely for technical reasons, as the
numerical solutions are easier to obtain with all boundary conditions
at one side, $w=y_{0}^{2}$\,].
The qualitative behavior of the $\sigma(w)$ curves
in these figures changes: the $\sigma(w)$ curves of
Figs.~\ref{fig3} and \ref{fig4} approach unity
from above as $w\to\infty$
[resulting in $M_{\mathsf{ADM}}>0$, according to
\eqref{eq:def-l} and \eqref{eq:M-ADM}],
while the $\sigma(w)$ curve of Fig.~\ref{fig5} approaches unity
from below as $w\to\infty$ [resulting in $M_{\mathsf{ADM}}<0$].

\section{Antigravity}
\label{sec:Antigravity}

\subsection{Origin of a new phenomenon}
\label{subsec:Origin-of-a-new-phenomenon}

Any localized object made of ponderable matter
(e.g., quarks and leptons of the Standard Model)
\textbf{attracts} a distant test particle.
This phenomenon is called \textbf{gravity} and
was first studied by Newton in his \textit{Principia} [1687].

With the solution of Fig.~\ref{fig5},
we have a localized object which \textbf{repels}
a distant test particle.
The phenomenon may be called ``\textbf{antigravity}.''
The crucial ingredients of this particular
object are, in the framework of Einstein's
General Relativity [1915], the nontrivial topology of space
[here, $\mathbb{R}P^3$\,]
and the nontrivial gravitational fields
at the defect surface [here, $\sigma(b^{2})<1$ for the
\textit{Ansatz} \eqref{eq:metric-Ansatz}].

Figures~\ref{fig4} and \ref{fig5} suggest
that the Skyrmion spacetime defect can have
either positive or negative gravitational mass,
but we have a further result:
\mbox{\emph{a sufficiently small defect solution exists}}
\emph{only if it has negative gravitational mass}
[a more precise formulation will be given later].

Consider, first, the nature of the solutions with
``standard'' boundary condition $\sigma(b^{2})=1$.
It is, then, found that the solution collapses if
it becomes too small
(loosely speaking, if $b$ is of the order of or less than
the effective Schwarzschild radius);
see Fig.~\ref{fig6} for numerical results at one particular
value of the coupling constant $\widetilde{\eta}$.
In fact, there is a \emph{critical curve}
in the $(b,\,\widetilde{\eta})$ plane, above which
(or, to the left of which)
there are no globally regular solutions with $\sigma(b^{2})=1$;
see Fig.~\ref{fig7} for a numerical approximation of this critical curve.

Let us now discuss the heuristics for obtaining
an anti-gravitating spacetime defect.
In order to get a globally regular solution in the
region \emph{above} the critical curve of Fig.~\ref{fig7},
we need to arrange for a sufficiently negative effective mass at
the defect surface ($y=0$).
Now, the dimen\-sion\-less effective mass is given by \eqref{eq:def-l}
with $w \equiv y_{0}^{2}+y^{2}$.
Then, an effective mass $l(y_{0}^{2})<0$ results from 
a nonstandard boundary condition $\sigma(y_{0}^{2})<1$.

For a fixed positive value
of $\widetilde{\eta}\equiv 8\pi\, G_{N}\, f^{2}$,
a sufficiently small globally
regular defect solution thus requires
a sufficiently negative effective mass at the
defect surface $w=y_{0}^{2}$  from a nonstandard boundary condition on 
the \textit{Ansatz} function for the $yy$ component of the metric, namely,
a positive value of $\sigma(y_{0}^{2})$ sufficiently far below unity.
For a small enough value of the
coupling constant $\widetilde{\eta}$,
this boundary condition at the defect surface directly gives
a negative ADM mass at spatial infinity
[see \eqref{eq:l-ODE} with a near-zero right-hand-side from
$\widetilde{\eta}\sim 0$
and the definition \eqref{eq:M-ADM} of $M_\text{ADM}$].

\subsection{Antigravity from a Planck-scale defect}
\label{subsec:Antigravity-from-a-Planck-scale-defect}

Let us give an explicit example~\cite{KlinkhamerQueiruga2018-antigrav}
of a negative-gravitational-mass Skyrmion spacetime defect.
With $\zeta$ a number of order 1 or larger,
the \text{model parameters} are taken as follows:
\begin{subequations}\label{eq:matter-theory-f2bound-ebound}
\begin{eqnarray}\label{eq:matter-theory-f2bound}
f^{2}  &\ll&    \big(E_\mathsf{planck}\big)^{2}
     \equiv   1/(8\pi G_{N})
     \approx  \big(2.44\times 10^{18}\;\mathsf{GeV}\big)^{2}\,,
\\[3mm]
\label{eq:matter-theory-ebound}
e    &\leq& 1/\zeta\,,
\end{eqnarray}
\end{subequations}
where the first inequality corresponds
to $\widetilde{\eta} = \big(f/E_\mathsf{planck}\big)^{2} \ll 1$.
The defect is considered to be\newpage 
\noindent 
a remnant of a quantum-spacetime phase and the defect length scale
is given by
\begin{subequations}\label{eq:b-remnant-l-planck}
\begin{eqnarray}
\label{eq:b-remnant}
b_\mathsf{remnant} &=& \zeta\;l_\mathsf{planck}\,,
\\[2mm]
\label{eq:l-planck}
l_\mathsf{planck}
             &\equiv& \sqrt{8\pi G_{N}\, \hbar/c^3}
             \approx 8.10 \times 10^{-35}\;\mathsf{m}\,,
\end{eqnarray}
\end{subequations}
with $\zeta \gtrsim 1$. The quantum nature of $b_\mathsf{remnant}$
has been emphasized by displaying $\hbar$ in \eqref{eq:l-planck}.

\begin{figure}[t]
\begin{center}
\includegraphics[width=0.65\linewidth]{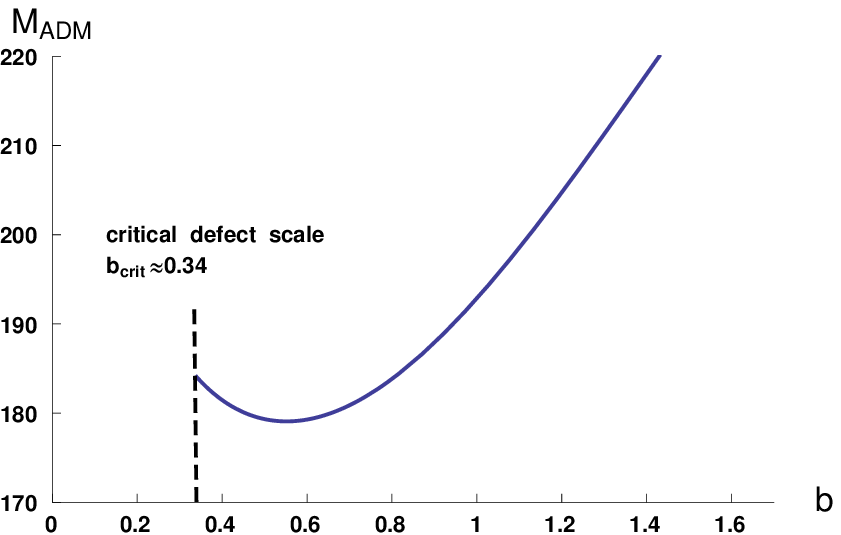}
\end{center}
\caption{Gravitational mass $M_\mathsf{ADM}$
[in units $f/e$] vs. defect length scale $b$ [in units $1/(ef)$]
from the numerical solutions~\cite{KlinkhamerQueiruga2018-antigrav}
of the ODEs \eqref{eq:ODEs-with-pion-mass}
with boundary condition $\sigma(b^{2})=1$,
for model parameters $\widetilde{\eta}=0.033$ and $\widetilde{m}=0$.}
\label{fig6}
\vspace*{0.125\baselineskip}
\begin{center}
\includegraphics[width=0.65\linewidth]{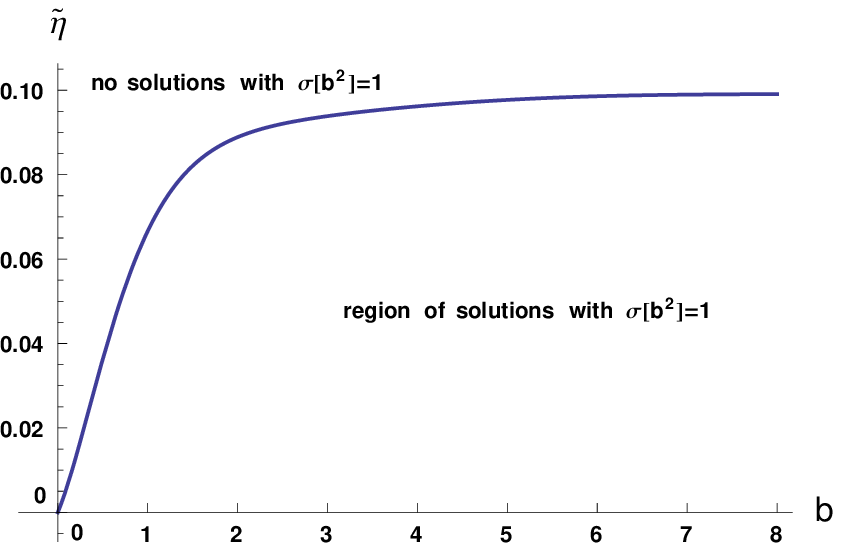}
\end{center}
\caption{Curve of the critical defect length scale $b_\mathsf{crit}$
[in units of $1/(e f)$] and the corresponding critical coupling constant
$\widetilde{\eta}_\mathsf{crit}$, obtained from
the numerical solutions~\cite{KlinkhamerQueiruga2018-antigrav}
of the ODEs \eqref{eq:ODEs-with-pion-mass} with boundary
condition $\sigma(b^{2})=1$ and model parameter $\widetilde{m}=0$.}
\label{fig7}
\end{figure}

The bounds \eqref{eq:matter-theory-f2bound-ebound}
and the defect length scale \eqref{eq:b-remnant-l-planck} imply
that the solution corresponds to a point
above the critical curve of Fig.~\ref{fig7},
so that a negative effective mass at the defect surface
is required in order to prevent collapse. 
The resulting gravitational mass of the defect 
[obtained from (\ref{eq:M-ADM}) with $l_\infty \sim -1$] is 
\begin{equation}\label{eq:negative-M-ADM--planck}
M_\mathsf{ADM} \sim
- \frac{4\pi}{\widetilde{\eta}}\,\frac{f}{e}
=
- \left(\frac{4\pi}{e}\right)\,
  \left(\frac{E_\mathsf{planck}}{f}\right)\,
  E_\mathsf{planck}\,.
\end{equation}
According to the inequalities \eqref{eq:matter-theory-f2bound}
and \eqref{eq:matter-theory-ebound}, the absolute value
$|M_\mathsf{ADM}|$ from \eqref{eq:negative-M-ADM--planck} typically
is much larger than $E_\mathsf{planck} \approx 4.34 \times10^{-6}\,\text{g}$.

\section{Stealth defect}
\label{sec:Stealth-defect}

\begin{figure*}[b] 
\vspace*{1mm}
\begin{center} 
\includegraphics[width=0.8\textwidth]{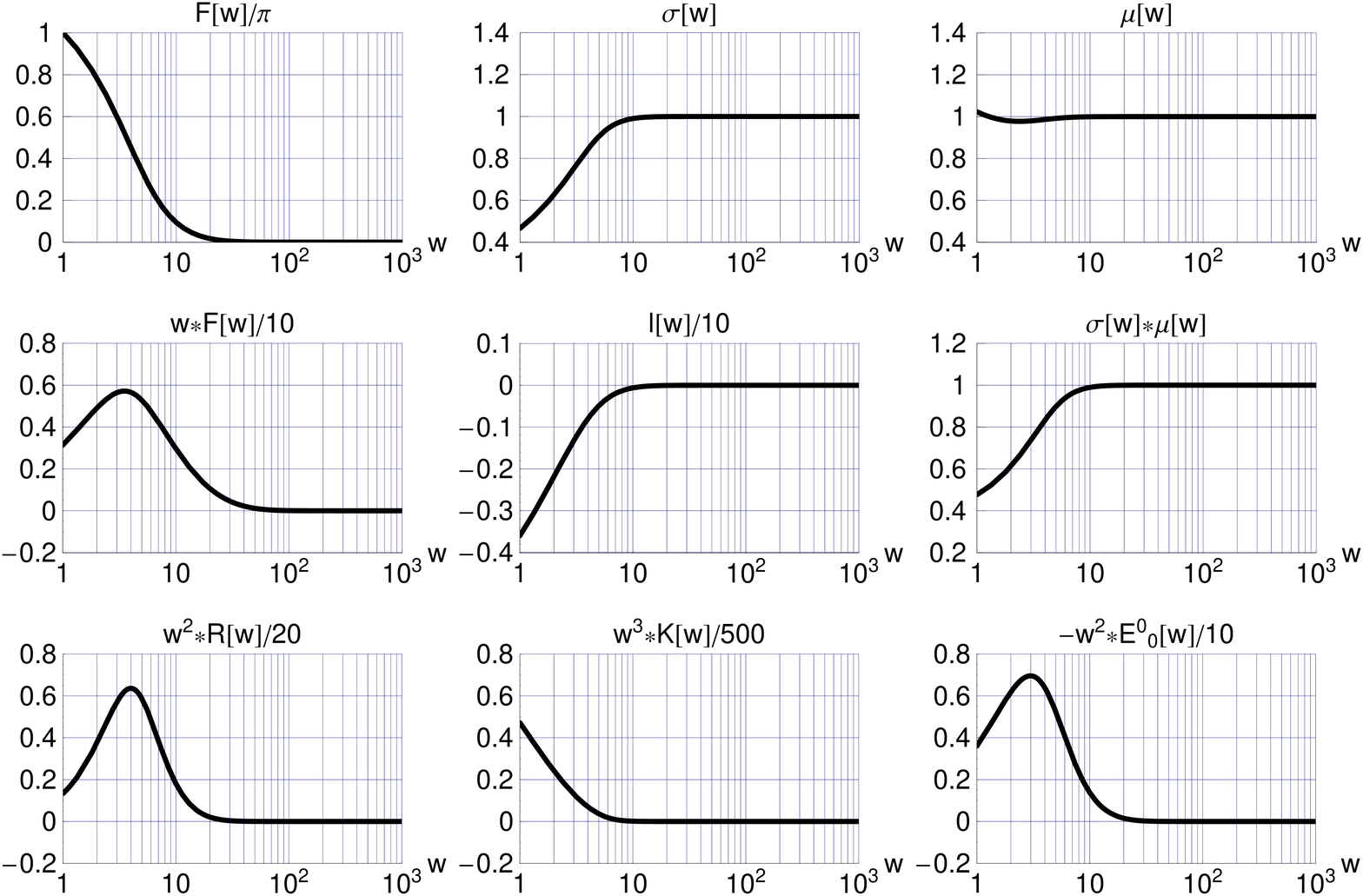}
\end{center}
\vspace*{-4mm}
\caption{Numerical solution~\cite{KlinkhamerQueiruga2018-stealth}
of the ODEs \eqref{eq:ODEs-with-pion-mass}
with parameters
$\widetilde{\eta}=1/10$, $\widetilde{m}=1$,  and $y_{0} =1$.
The boundary conditions at $w=1$ are:
$F/\pi=     1.00000$,
$F^\prime=  -0.752388$,
$\sigma=    0.466343$,
and  $\mu=  1.02282$.
The value of $|l(10^3)|$ is less than $10^{-11}$, which gives
an essentially vanishing ADM mass \eqref{eq:M-ADM}.
The matter energy density $\rho(w)$ is proportional to 
$-E^{0}_{\;\;0}(w)$ by the Einstein equation.
The energy density $\rho(w)$ of this numerical solution 
vanishes exponentially as $w \to \infty$.
}
\label{fig8}
\vspace*{-1mm}  
\end{figure*}

We have seen
that certain soliton-type defect solutions can have positive
gravitational mass but also negative gravitational mass.
As the gravitational mass of such a spacetime-defect solution
is a continuous variable, there must be
special spacetime defects with vanishing gravitational mass.
These defects with positive energy density of
the matter fields and zero asymptotic gravitational mass will be called
``stealth defects''~\cite{KlinkhamerQueiruga2018-stealth}.
An explicit solution with vanishing gravitational
mass and an exponentially-vanishing
energy density of the matter fields
is given in Fig.~\ref{fig8}.

Now, assume that all matter fields have some form of
non-gravitational interaction with each other.
If so, there will be some interaction
between the ``pions'' of the theory considered in
(\ref{eq:action-Lgrav-Lmat-omegamu}) and the elementary
particles of the Standard Model.
Then, consider what happens with the head-on collision
of a stealth defect from Fig.~\ref{fig8} and a human observer made of
Standard Model particles
(mostly up and down quarks, gluons, and electrons).
In close approximation, the observer will have no idea
of what is going to happen, until he/she is within
a distance of order $\hbar/(m c)$ from the defect, where
$m$ is the ``pion'' mass scale from the
matter action (\ref{eq:Lmat}).
What happens during the collision itself
and afterwards depends on the details of the setup,
for example, the size of the observer compared to the
defect length scale $b$.

\section{Lensing by a stealth defect}
\label{sec:Lensing}

Consider the behavior of light rays propagating over a spacetime-defect
manifold. It is, in fact, possible to
give a simplified discussion~\cite{KlinkhamerWang2019-lensing}
by use of an exact vacuum solution~\cite{Klinkhamer2014-mpla},
\begin{subequations}\label{eq:vacuum-metric-w-def-y0-def}
\begin{eqnarray}\label{eq:vacuum-metric-general-l}
\hspace*{-5mm}
ds^{2}\,\Big|^\mathsf{(vac.\,sol.)}_{\widetilde{M}_4\,,\,\mathsf{chart-2}}
&=&
-\left(1-\widehat{l}/\sqrt{w}\,\right)\,(dt)^{2}
+\frac{1-y_{0}^{2}/w}{1-\widehat{l}/\sqrt{w}}\;(dy)^{2}
+w\,\Big[(dz)^{2}+\sin^{2}z\,(dx)^{2}\Big]\,,
\\[1mm]
\label{eq:w-def}
\hspace*{-5mm}
w &\equiv& y_{0}^{2} +y^{2}\,,
\quad
y_{0} \equiv e\, f\, b  > 0\,,
\quad
\widehat{l} \in (-\infty,\, y_{0})\,.
\end{eqnarray}\end{subequations}
In the notation of (\ref{eq:metric-Ansatz}) with dimensionless
variables, we have
$[\mu(w)]^{2}=1/[\sigma(w)]^{2}=1-\widehat{l}/\sqrt{w}\,$.
In our setup, the vacuum metric \eqref{eq:vacuum-metric-w-def-y0-def}
can arise in two ways: first, from the solution of
Sec.~\ref{sec:Skyrmion-spacetime-defect} with nontrivial
matter-field boundary conditions \eqref{eq:bcs-F}
in the limit $G_N \to 0$ for fixed energy scale $f>0$ or, second,
from the metric of Sec.~\ref{sec:Skyrmion-spacetime-defect}
with a trivial matter field $F(W)=0$, so that the matter energy
density \eqref{eq:T-up0-down0} vanishes identically.

Next, consider the special case of a stealth-defect vacuum solution 
\eqref{eq:vacuum-metric-w-def-y0-def} with
\begin{equation}\label{eq:widehat-l-zero-case}
\widehat{l}=0\,,
\end{equation}
which has a flat spacetime with vanishing curvature scalars,
$R(w)=K(w)=0$.
We consider this ``extreme'' case, in order to emphasize
the difference with standard gravitational lensing~\cite{Schneider-etal1992}
which is due to the curvature of spacetime resulting from a nonvanishing
matter distribution.

Remark that  \emph{exact} multi-defect solutions
of the vacuum Einstein equation can be obtained by
superposition of these static $\widehat{l}=0$ defects,
provided the individual defect surfaces do not intersect.
The resulting ``gas'' of static defects violates Lorentz invariance
(see Sec.~6 in Ref.~\cite{KlinkhamerWang2019-lensing} for further details).
In principle, we can also obtain an exact
multi-defect solution of the vacuum Einstein equation
which is approximately Lorentz invariant, if we superpose
quasi-randomly positioned and quasi-randomly moving
$\widehat{l}=0$ defects (arranged to be nonintersecting initially).

\begin{figure}[p!]
\centering
\includegraphics[width=0.6\textwidth]{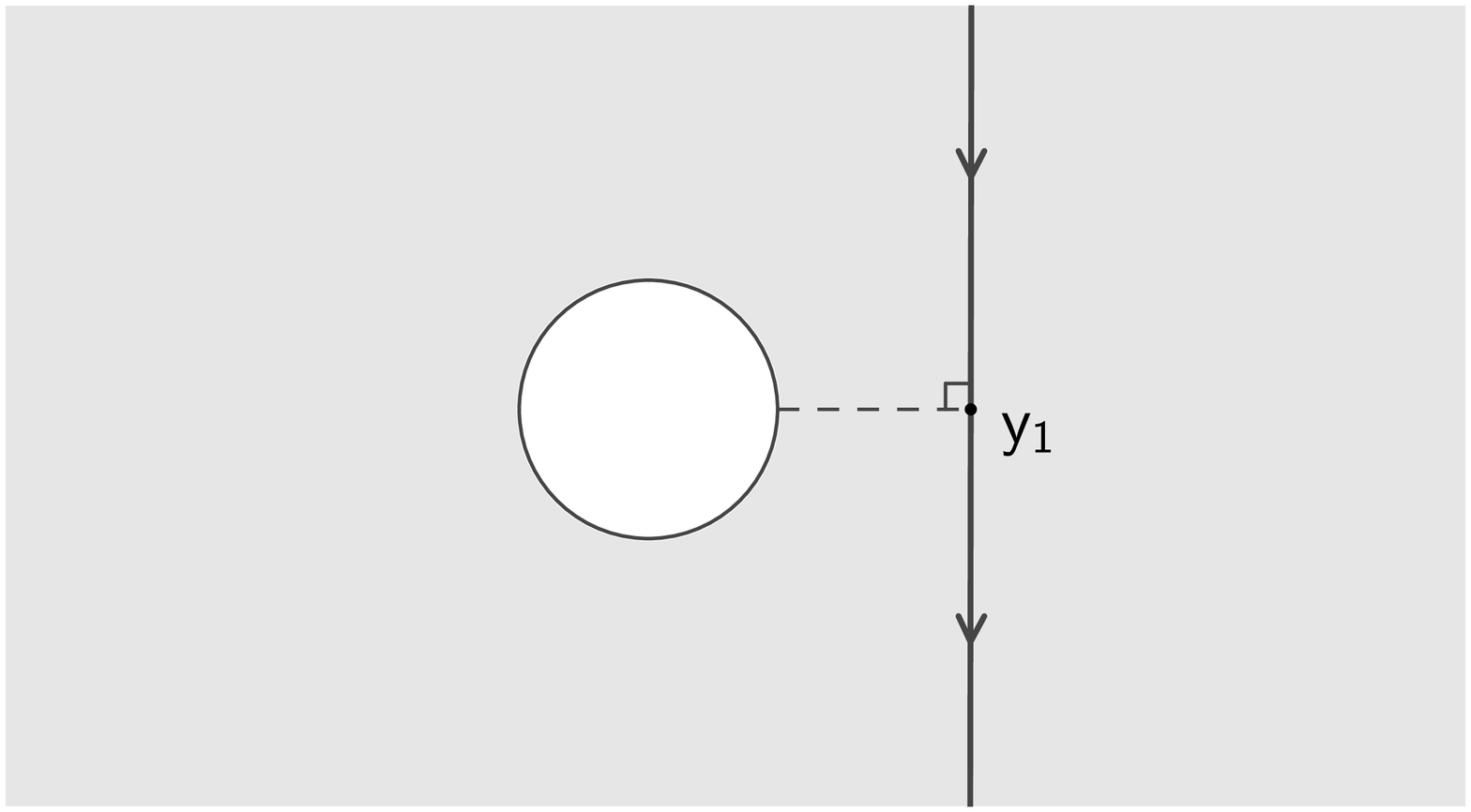}
\caption{Geodesic which does not cross the defect surface,
with part of the 3-space manifold
\eqref{eq:vacuum-metric-w-def-y0-def}--\eqref{eq:widehat-l-zero-case}
indicated by the shaded area.
The dimensionless quasi-radial coordinate $y_1$ corresponds to an
``impact parameter.''}
\label{fig9}
\vspace*{2\baselineskip}
\vspace*{5mm}
\centering
\includegraphics[width=0.6\textwidth]{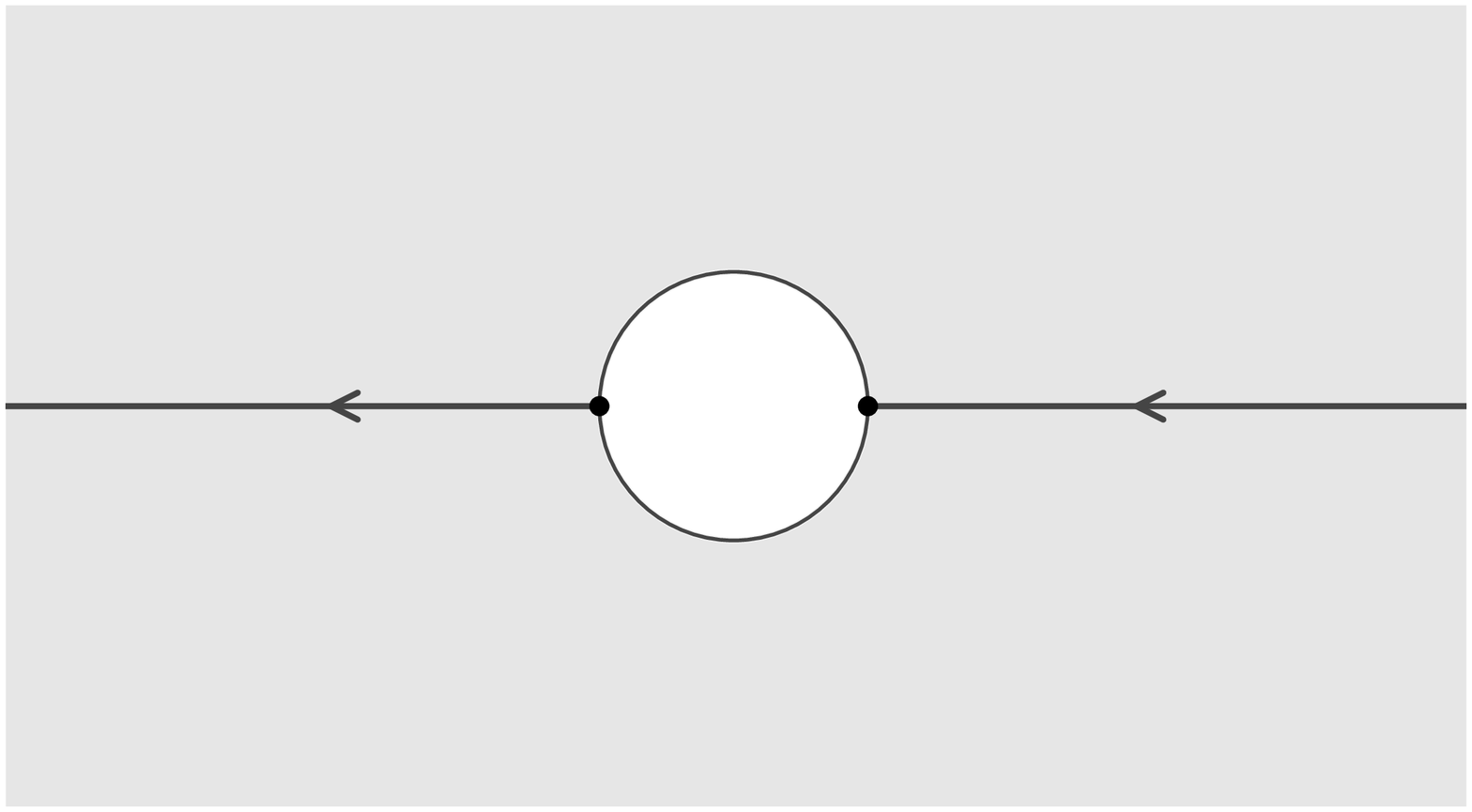}
\caption{Radial geodesic which   crosses the defect surface, where
antipodal points (dots) on the defect surface are identified
(cf. Fig.~\ref{fig1}).}
\label{fig10}
\vspace*{2\baselineskip}
\vspace*{5mm}
\centering
\includegraphics[width=0.6\textwidth]{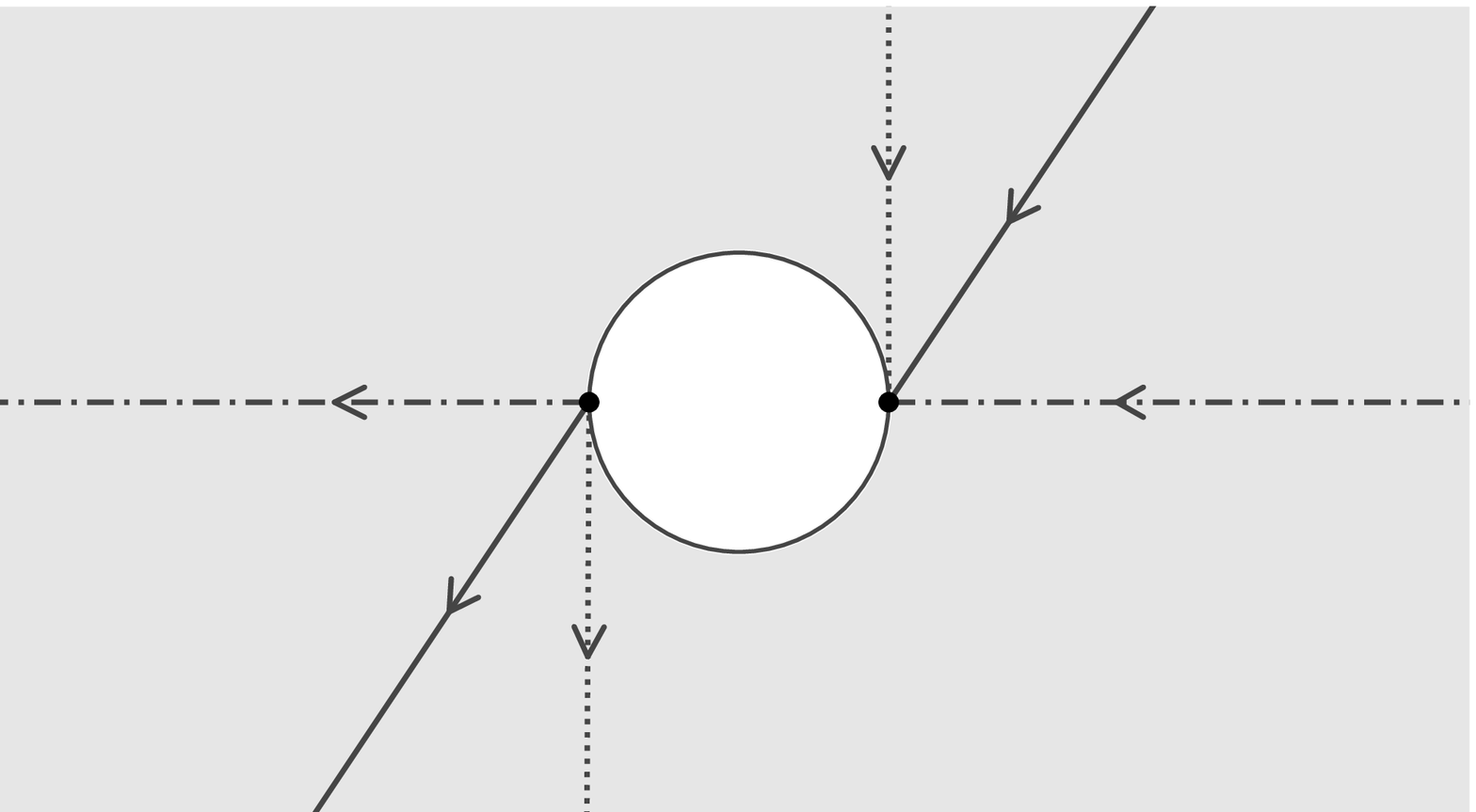}
\caption{A family of geodesics crossing the defect surface.}
\vspace*{-10mm}  
\label{fig11}
\end{figure}

Returning to the single $\widehat{l}=0$
defect with metric \eqref{eq:vacuum-metric-w-def-y0-def}, 
the geodesics are readily calculated:%
\begin{itemize}
\item
straight lines in the ambient Euclidean 3-space,
if there are no intersections with the defect surface;
  \item
matching straight-line segments in the ambient Euclidean 3-space,
if there are intersections with the defect surface.
\end{itemize}
Figure~\ref{fig9} gives an example of a geodesic staying away from
the defect surface, while Figs. \ref{fig10} and \ref{fig11}
show geodesics crossing the defect surface,
with or without parallel shift of the straight-line segments
in the ambient Euclidean 3-space
(the solid line in Fig.~\ref{fig11}, for example, has a parallel shift
but the dot-dashed line not).

\begin{figure}[t!]
\centering
\includegraphics[width=0.6\textwidth]{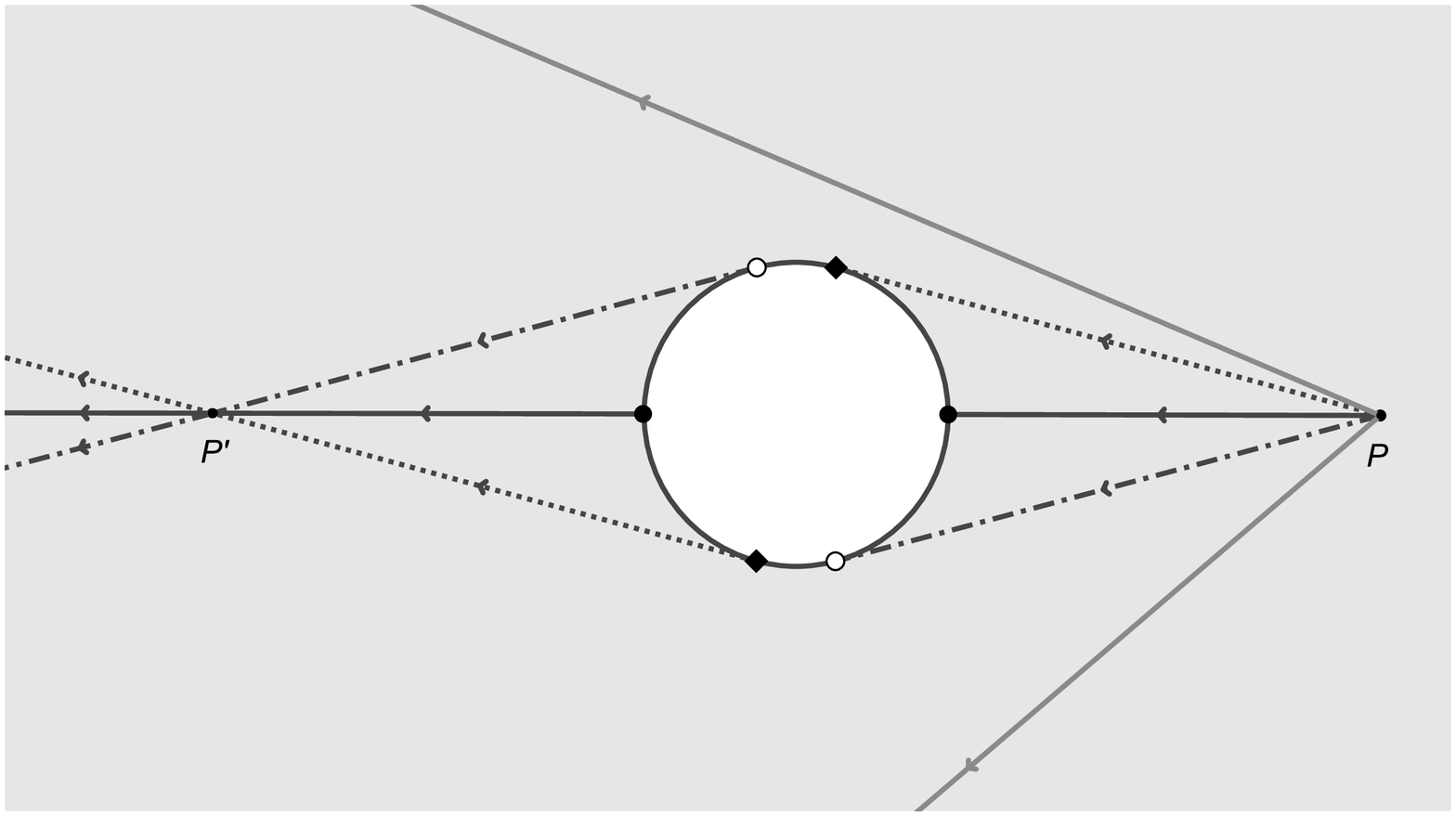}
\caption{Geodesics with intersection points $P$ and $P'$.
}
\label{fig12}
\vspace*{2\baselineskip}
\centering
\includegraphics[width=0.6\textwidth]{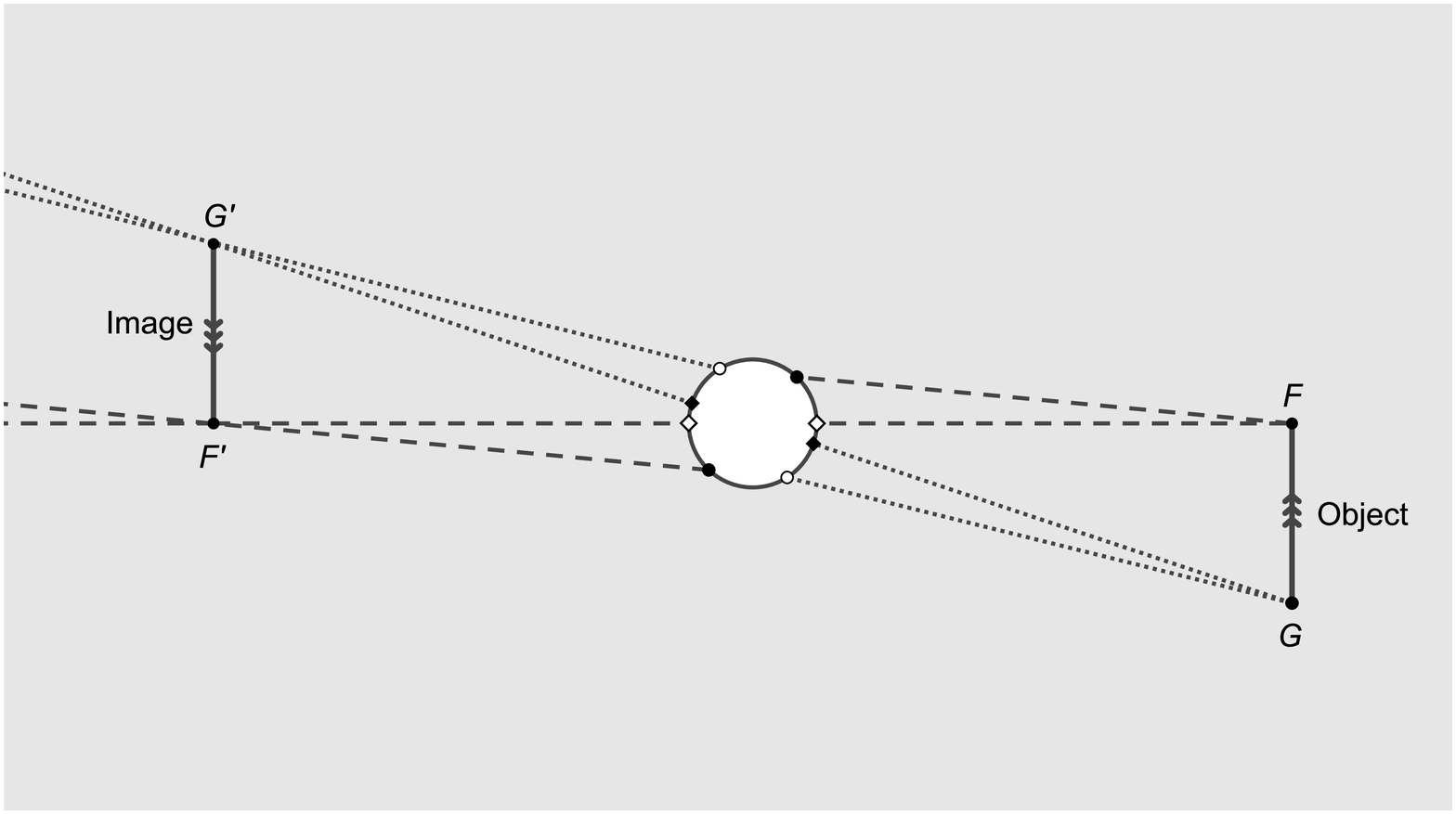}
\caption{Image formation by the stealth defect.
}
\label{fig13}
\vspace*{-2mm}
\end{figure}

Due to the parallel shifts at the defect surface,
there is a lensing effect, as shown in Fig.~\ref{fig12}.
This lensing of the flat-spacetime defect results in image formation,
as illustrated by  Fig.~\ref{fig13}.
A few remarks are in order:
\begin{enumerate}
  \item
The image in Fig.~\ref{fig13} is located at the reflection
point on the other side of the defect.
  \item
The image is inverted and the image size is equal to the object size.
Note that this is also the case if an object in Minkowski spacetime
is located at a $2f$ distance from a thin double-convex lens,
where $f$ is the focal length of the lens.
  \item
The irradiance of the image (defined as the power per unit receiving area)
depends on the defect length scale $b$ and the location of the object:
the irradiance of the image will be larger
if $b$ is increased for unchanged object position
or if the object is brought closer to the defect for unchanged $b$.
  \item
If a permanent pointlike light source is placed
at point $P$ of Fig.~\ref{fig12},
then an observer at point $P'$ in the same figure will see
a \emph{luminous disk}
(different from the \emph{Einstein ring}~\cite{Schneider-etal1992}
which the observer would see if the defect were replaced
by a patch of Minkowski spacetime with a static
spherical star at the center).
\end{enumerate}

\section{Discussion}
\label{sec:Discussion}

Before we review the main characteristics
of our soliton-type spacetime defect,
we wish to compare our approach to another approach which is
more or less orthogonal.\vspace*{-2mm}

\subsection{Comparison to another kind of spacetime defect}
\label{subsec:Comparison-to-another-kind-of-spacetime-defect}

It is, in principle, possible to define a local spacetime defect
not by what it \emph{is} but rather by what it \emph{does}.
Precisely this approach has been adopted in
Refs.~\cite{Hossenfelder2013,Hossenfelder-etal2018},
where the effect considered is the violation of energy-momentum
conservation.
More precisely, a single particle moves in the ambient spacetime
(i.e., the smooth spacetime away from the localized defect) and
is arranged to ``collide'' with a pointlike spacetime defect,
which randomly changes the 4-momentum of the particle.
The problem with this approach is how to implement it into a consistent theory,
without knowing the details of the defect.

In the flat-spacetime model of Ref.~\cite{Hossenfelder2013},
the suggestion is to modify the gauge-covariant
derivative appearing in the Lagrange density.
But, then, gauge invariance is violated,
which will introduce an unacceptable violation
of unitarity for the particles in the ambient spacetime.

In the curved-spacetime model of Ref.~\cite{Hossenfelder-etal2018},
the standard Levi-Civita connection is modified by the addition
of a rank-3 tensor $Q^{\lambda}_{\mu\nu}=Q^{\lambda}_{\nu\mu}$,
which is thought to be defined on a discrete set of spacetime points
(corresponding to the pointlike spacetime defects).
The problem, now, is to obtain the modified Einstein equation,
which somehow involves derivatives of $Q^{\lambda}_{\mu\nu}$,
even though these $Q^{\lambda}_{\mu\nu}$ are only defined
on a discrete set of points. It is perhaps
possible to consider some smeared-out version of $Q^{\lambda}_{\mu\nu}$,
but this is not really satisfactory for a fundamental theory.

Another possibility is simply to view
the model of Ref.~\cite{Hossenfelder-etal2018} 
as an effective theory over a smooth manifold, with
pointlike defects replaced by finite-size soliton-type defects.
Instead of static soliton-type defects, we then need
genuine time-dependent solutions of the classical field equations.
Alternatively, we may consider our Skyrmion spacetime defect
as a first approximation, where the
defect size is considered to be time-dependent:
$b=b(t)$ with an approximately constant value
$b(t)\sim b_{c}>0$ for $|t| < \Delta t$ and
an approximately vanishing value $b(t)\sim 0^{+}$ for $|t| > \Delta t$.
This last suggestion for $b(t)$ would correspond to
``topology change without topology change''
(cf. Sec.~6.6.4 in Ref.~\cite{Visser1995}).
The issue of topology change will also be discussed
at the end of Sec.~\ref{subsec:Remarks-on-the -Skyrmion-spacetime-defect}.

Expanding on the last paragraph, we remark that it is
actually possible to calculate the scattering of a ``pion''
[as defined in \eqref{eq:Omega-pi-a}] by the $SO(3)$ Skyrmion
spacetime defect and to obtain the recoil of the defect
with the corresponding  energy-momentum change of the ``pion.''
The main observation is that our Skyrmion spacetime defect
has a finite-width shell of nonvanishing pion-density,
as follows from setting $F(w) \sim \pi$
in \eqref{eq:T-up0-down0} for $w$ values between $y_{0}^{2}$
and  $y_{0}^{2}+(\Delta y)^{2}$.
This finite matter-density shell then scatters a
single incoming ``pion.''
In fact, pion-nucleon scattering~\cite{Hayashi-etal1984,MattisPeskin1985}
has been discussed for the standard $SU(2)$ Skyrmion in Minkowski spacetime
and the recoil of this Skyrmion  (a finite matter-density ball) has been
obtained in Ref.~\cite{Uehara-etal1991,HughesMathews1992}.
A similar calculation of the recoil would appear to be feasible for
our $SO(3)$ Skyrmion spacetime defect
(a finite matter-density shell).
This calculated recoil would provide an explicit
realization for the defect-induced
energy-momentum change~\cite{Hossenfelder-etal2018} of the incoming
particle (here, a ``pion'').

\subsection{Remarks on the Skyrmion spacetime defect}
\label{subsec:Remarks-on-the -Skyrmion-spacetime-defect}

The following general remarks aim to put our Skyrmion spacetime defect
solution in perspective.
First, the new type of Skyrmion solution is rather interesting by itself,
as it combines the nontrivial topology of spacetime with the
nontrivial topology of field-configuration space.
Indeed, the nontrivial topology of the underlying space manifold
allows the internal $SO(3)$ space to be covered only once,
$N=1$ in \eqref{eq:deg-Omega}.

Second, it remains to be proved that the solution obtained is stable.
The scalar fields by themselves would be stable because of the
topological charge $N=1$, but there could still be more
branches of solutions with even lower values of the ADM mass
(the two known branches are shown in Fig.~2 of
Ref.~\cite{KlinkhamerQueiruga2018-antigrav}).

Third, the Skyrmion-spacetime-defect metric
from (\ref{eq:Ansaetze})
is degenerate: $\mathsf{det}\, g_{\mu\nu}=0$ at the defect surface $Y = 0$,
which corresponds to a submanifold $\mathbb{R}P^{2} \sim S^{2}/\mathbb{Z}_2$
(cf. Fig.~\ref{fig1}).
A heuristic argument for the necessity of a degenerate
metric is as follows:
the submanifold $\mathbb{R}P^{2}$ cannot be differentially embedded
in $\mathbb{R}^3$ and this fact implies that
a nonsingular solution requires a vanishing
metric component $g_{YY}$ at $Y=0$, making for a vanishing
determinant at $Y=0$ (more details can be found in the third remark
of Sec.~VI in Ref.~\cite{KlinkhamerQueiruga2018-antigrav}).

Fourth, this degenerate metric makes that the Gannon singularity
theorem~\cite{Gannon1975}
and the Schoen--Yau positive-mass theorem~\cite{SchoenYau1979}
are not directly applicable; we refer to Sec.~3.1.5 in
Ref.~\cite{Guenther2017} for further discussion.
The special feature of the Skyrmion spacetime
defect solution is that
certain geodesics at the $\mathbb{R}P^{2}$
defect surface cannot be continued
uniquely: compare the full curve of Fig.~\ref{fig9}
in the limit of $y_1 \to 0^{+}$
with the dotted curve of Fig.~\ref{fig11}.

Fifth, the negative ADM mass found
for small enough defect length scale $b$
(at a given value of $\widetilde{\eta}$)
is not due to ponderable matter but to nontrivial
gravitational fields at the $\mathbb{R}P^{2}$ defect
surface with area $2\pi b^{2}$.
Specifically, the boundary value $\sigma(b^2)\in (0,\,1)$
determines the approach to zero of the metric
component $g_{YY}$, according to the metric
\textit{Ansatz} \eqref{eq:metric-Ansatz}.

Sixth, having a negative-gravitational-mass spacetime defect
[notably a vacuum solution \eqref{eq:vacuum-metric-general-l} with
$\widehat{l}<0$] prompts us to reconsider the
stability of Minkowski spacetime, especially as classical
topology change appears to be possible if degenerate metrics
are allowed~\cite{Horowitz1991}.

Seventh, the crucial open question
is the origin and role of nontrivial spacetime topology.
For our Skyrmion spacetime defect solution, two specific questions are
\begin{enumerate}
\item
what sets the constant defect length scale $b$?
\item
can the defect length scale $b$ become a dynamic variable?
\end{enumerate}
There are, of course, many further questions, such as
what happens if two defects collide?

The two specific questions listed in the last remark
have also been raised in the penultimate paragraph of
Sec.~\ref{subsec:Comparison-to-another-kind-of-spacetime-defect}.
The issue of (genuine or effective)
topology change is essential for a proper understanding
of the small-scale structure of spacetime, which
brings us back to the ``quantum phase''
of the first sentence in Sec.~\ref{sec:Introduction}.

\ack
The author thanks the organizers
of this interesting workshop, in particular Hans-Thomas Elze.
He also thanks J.M. Queiruga and Z.L. Wang for
useful comments on the manuscript.

\end{document}